\documentclass[12pt]{article}
\usepackage{amssymb,graphicx}
\usepackage{epstopdf}
\usepackage{amsmath,amsfonts}
\usepackage{epsfig}

\def\d{\partial}
\def\l{\left(}
\def\r{\right)}

\newcommand{\be}{\begin{equation}}
\newcommand{\ee}{\end{equation}}
\newcommand{\bea}{\begin{eqnarray}}
\newcommand{\eea}{\end{eqnarray}}
\newcommand{\bg}{\begin{gather}}
\newcommand{\eg}{\end{gather}}
\newcommand{\bseq}{\begin{subequations}}
\newcommand{\eseq}{\end{subequations}}

\begin{document}
\begin{flushright}
INR-TH/2014-020\\
CERN-PH-TH/2014-181
\end{flushright}

\vspace{10pt}
\begin{center}
  {\LARGE \bf On holography for \\[0.3cm] (pseudo-)conformal cosmology } \\
\vspace{20pt}
M.~Libanov$^{a,b}$, V.~Rubakov$^{a,c}$, S.~Sibiryakov$^{a,d,e}$\\
\vspace{15pt}

$^a$\textit{
Institute for Nuclear Research of
         the Russian Academy of Sciences,\\  60th October Anniversary
  Prospect, 7a, 117312 Moscow, Russia}\\
\vspace{5pt}

$^b$\textit{Moscow Institute of Physics and Technology,\\
Institutskii per., 9, 141700, Dolgoprudny, Moscow Region, Russia
}

\vspace{5pt}

$^c$\textit{Department of Particle Physics and Cosmology,
Physics Faculty, Moscow State University\\ Vorobjevy Gory,
119991, Moscow, Russia}

\vspace{5pt}

$^d$\textit{Theory Group, Physics Department, CERN, CH-1211 Geneva 23,
Switzerland}

\vspace{5pt}

$^e$\textit{FSB/ITP/LPPC,
Ecole Polytechnique Federale de Lausanne, CH-1015 Lausanne, Switzerland
}
    \end{center}
    \vspace{5pt}

\begin{abstract}

We propose a holographic dual for
(pseudo-)conformal cosmological scenario, with a scalar field
that forms a moving domain wall in adS$_5$. The domain wall
separates two vacua with unequal
energy densities. Unlike in the existing construction, the 5d solution
is regular in the relevant space-time domain.

\end{abstract}

It has been understood for some time that the (nearly) flat spectrum
of scalar cosmological perturbations may be a consequence of
conformal symmetry $SO(4,2)$ broken down to de~Sitter $SO(4,1)$ in the
early Universe~\cite{Rubakov:2009np,Creminelli:2010ba,Hinterbichler:2011qk}.
This has lead to (pseudo-)conformal cosmological scenario which serves as an
alternative to inflation. In general terms~\cite{Hinterbichler:2012mv},
the symmetry breaking pattern is realized when a scalar operator
(or several operators)
${\cal O}$ of non-zero conformal weight $\Delta$ acquires the time-dependent
expectation value
\be
\langle {\cal O} \rangle = \frac{\mbox{const}}{(-t)^{\Delta}}\; ,
\label{sep15-14-1}
\ee
where $t<0$ and the rolling regime is supposed to terminate at
some finite negative $t$, cf.~Ref.~\cite{Wang:2012bq}. Other ingredients
of the scenario are: (i) nearly flat space-time at the rolling stage
\eqref{sep15-14-1}, hence no primordial tensor perturbations;
(ii) another scalar field of zero conformal weight, whose perturbations
automatically have flat power spectrum at late times at the rolling
stage\footnote{Slight
explicit breaking of conformal invariance yields small tilt in this
spectrum~\cite{Osipov:2010ee}.}; (iii) conversion of these field
perturbations into adiabatic perturbations at some later epoch.
Among potentially observable predictions of this scenario are
statistical anisotropy of scalar
perturbations~\cite{Libanov:2010nk,Libanov:2011hh,Creminelli:2012qr}
and non-Gaussianities of specific
shapes~\cite{Creminelli:2012qr,Libanov:2011bk,Mironov:2013bza}.
These features are to large extent generic consequences of
the symmetry breaking pattern~\cite{Creminelli:2012qr}
 $SO(4,2) \to SO(4,1)$.

It is of interest to construct a holographic dual to
the (pseudo-)conformal scenario. A construction of this sort
has been proposed in
Ref.~\cite{Hinterbichler:2014tka}. It
employed free massive scalar field evolving in adS$_5$
with metric (hereafter the adS radius is set equal to 1)
\be
ds^2 = \frac{1}{z^2}\left(
\eta_{\mu \nu} dx^\mu dx^\nu - dz^2\right) \; .
\label{may10-2}
\ee
In the probe scalar field
approximation,
the corresponding solution is, however, singular at the null surface
$z=-t$; upon switching on 5d gravity, the solution develops
a naked singularity along the surface $z/(-t) = \mbox{const} < 1$.
This feature does not
look particularly desirable, so one is lead to search for
non-singular constructions.

On the other hand,
the (pseudo-)conformal scenario is naturally realized in a DBI
theory \cite{Hinterbichler:2012fr} descending from the dynamics of
a thin brane in adS$_5$ \cite{Goon:2011qf}. So, it is natural
to merge the two constructions and consider a thick brane
evolution in adS$_5$ as a candidate for the holographic dual to
the (pseudo-)conformal Universe. This is precisely the purpose of this
paper.

To this end, we adopt a bottom-up approach, and instead of constructing
a concrete CFT, simply
consider a 5d theory of a scalar field with action
\be
S = \int~\sqrt{g} \l \frac{1}{2} g^{AB} \d_A \phi \d_B \phi - V(\phi) \r ~
dz~d^dx
\ee
We will work in the probe scalar field approximation throughout,
so that the metric \eqref{may10-2} is unperturbed.
We would like this theory to correspond to a boundary CFT
without explicit breaking of conformal invariance, but with
unstable conformally invariant vacuum $\phi=0$. So, unlike in
Refs.~\cite{Distler:1998gb,Hertog:2004rz,Hertog:2005hu}
we assume that
the potential $V(\phi)$ has a local minimum at $\phi = 0$
and that the field
%
%
behavior near the adS boundary $z=0$
is
\be
\phi (z, x) = z^{\Delta_+} \phi_0 (x)
\label{dec10-14-1}
\ee
with $\Delta_+ = \sqrt{m^2+4} +2$, where $m$ is the scalar field mass in
the vacuum $\phi =0$, and $\phi_0$ is related to the expectation value of
a CFT operator \cite{Balasubramanian:1998de,Klebanov:1999tb},
\be
\langle {\cal O} \rangle = 2\sqrt{m^2 + 4} ~\phi_0 \; .
\label{dec10-14-3}
\ee
The property \eqref{dec10-14-1} implies that there is no explicit
deformation of the boundary CFT,
in contrast to Refs.~\cite{Hertog:2004rz,Hertog:2005hu,Craps:2007ch},
whereas we allow for spontaneous breaking of
conformal symmetry.
To avoid possible runaway behavior in 5d theory, the
potential $V(\phi)$ is assumed to have a global minimum
 at $\phi = \phi_+ \neq 0$. Finally, we assume that the curvature of
the potential at its maximum is large enough,
\be
|V^{\prime \prime}(\phi_{max})| > 2 \; ,
\label{dec10-14-2}
\ee
so that the solution we are about to discuss does not get stuck at
this maximum, see below. So, from our viewpoit the position of the
maximum is not particularly significant.

We are now going to construct a solution in adS for which
the expectation value of the operator ${\cal O}$ given by
\eqref{dec10-14-3} has the form \eqref{sep15-14-1}, i.e.,
$\langle {\cal O} \rangle \propto (-t)^{-\Delta_+}$. This is
a domain wall separating the vacua $\phi=0$ and $\phi=\phi_+$,
with the false vacuum $\phi=0$
to the left of it.

The symmetry breaking pattern
$SO(4,2) \to SO(4,1)$ is obtained
for a 5d solution of the form \cite{Hinterbichler:2014tka}
\be
\phi = \phi \left( \frac{t}{z} \right) \;
\label{sep12-14-1}
\ee
with asymptotics
\be
\phi (v) \to (-v)^{-\Delta_+} \;\;\;\; \mbox{as} \;\; \; v \to - \infty \; ,
\label{sep12-14-6}
\ee
where
\[
v = \frac{t}{z} \; .
\]
We require that the solution is non-singular at strictly negative $t$,
but not necessarily at $t \geq 0$,
since by assumption the rolling behavior \eqref{sep15-14-1} terminates
at some finite negative time. Formally, the ``point'' $t=0, z=0$ is special,
since it is reached from different directions, i.e., at different values of
$v$ and hence different values of $\phi$. This can be interpreted
as the ``point'' at which the wall hits
the adS boundary.

\begin{figure}[tb!]
\begin{center}
\includegraphics[width=0.6\textwidth,angle=0]{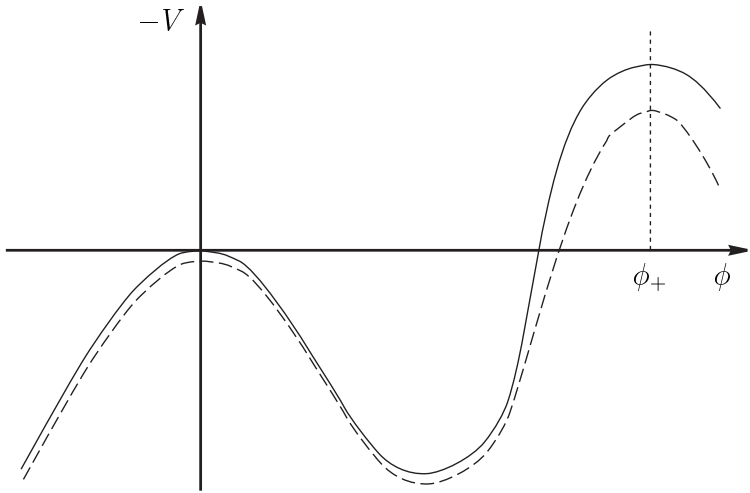}
\end{center}
\caption{Inverted potential $[-V(\phi)]$ (solid line)
and inverted auxiliary potential
$[-V_0 (\phi)]$ (dashed line).
\label{potential}
 }
 \end{figure}

With the Ansatz \eqref{sep12-14-1}, the field equation is
\be
(v^2 - 1)\frac{d^2 \phi}{d v^2} + 5v \frac{d \phi}{d v}
- \frac{\d V}{\d \phi} = 0 \; .
\label{sep12-14-2}
\ee
For a class of potentials $V(\phi)$ it does admit  a domain wall solution
which is non-singular at $v < 0$. To see this, consider first the
vicinity of a null surface $v = -1$; note that at this surface
eq.~\eqref{sep12-14-2} is singular
(the second derivative term vanishes). For any value of $\phi_* \equiv
\phi(v=-1)$, there exists a non-singular solution
\be
\phi = \phi_* - \frac{1}{5} V^\prime (\phi_*) \cdot (v+1)
-  \frac{1}{14} V^\prime (\phi_*) \left[ 1  - \frac{1}{5}
V^{\prime \prime} (\phi_*) \right] \cdot (v+1)^2 + \dots
\label{sep12-13-5}
\ee
For $v < -1$, we change the variable,
$v = - \cosh \rho$ (cf. Ref.~\cite{Hinterbichler:2014tka}) and write
eq.~\eqref{sep12-14-2} in the following form
\be
\frac{d^2 \phi}{d \rho^2} + 4\frac{\cosh \rho}{\sinh \rho} \cdot
\frac{d \phi}{d \rho}
- \frac{\d V}{\d \phi} = 0 \; .
\label{sep12-14-3}
\ee
This equation corresponds to a motion of a ``particle'' in the
inverted potential $-V$ with ``time''-dependent friction,
from $\rho =0$ ($z=-t$) to $\rho \to \infty$ ($z \to 0$).
Let $V_0 (\phi)$ be an auxiliary potential, such that there exists
a solution to
\be
\frac{d^2 \phi}{d y^2} + 4
\frac{d \phi}{d y}
- \frac{\d V_0}{\d \phi} = 0 \; ,
\label{sep15-14-5}
\ee
that starts at $y \to -\infty$ in the true vacuum
$\phi = \phi_+$ and reaches the false vacuum $\phi = 0$ at
$y \to \infty$. Note that the necessary condition for the
existence of such a solution is that $|V_0^{\prime \prime}| > 2$
at the maximum of $V_0$, cf. eq.~\eqref{dec10-14-2}.
Incidentally, this solution can be interpreted
as a static domain wall $\phi (y)$, where $y= - \log z$.
In our context, this would be a domain wall centered
at $\rho \to \infty$, i.e., asymptotically close to the adS
boundary, where eq.~\eqref{sep12-14-3} reduces to
eq.~\eqref{sep15-14-5}. Now, let the potential $V(\phi)$
be slightly deeper than
$V_0$ in the true vacuum, see Fig.~\ref{potential}.
Then, by continuity, there exists
a value of $\phi_* <  \phi_+$ such that
the solution to  eq.~\eqref{sep12-14-3} starts at
$\rho =0$ from
$\phi = \phi_*$ (at zero velocity $d \phi / d \rho$, since
$v = - 1 - O(\rho^2)$, see eq.~\eqref{sep12-13-5})
and approaches the false vacuum
$\phi =0$ as $\rho \to \infty$: a solution starting very close to
the top of $(-V)$ overshoots the true vacuum, while a solution
that starts from small $\phi$ undershoots it. So, at least for potentials
sufficiently close to $V_0$, there exists a
domain wall which is non-singular in the region $z< -t$.
It has the asymptotics \eqref{sep12-14-6}.

Since the solution is non-singular at $z=-t$, it can be continued to
the region $z>-t$. There, we change the variable to
\[
v = - \cos \theta
\]
and find from eq.~\eqref{sep12-14-2}
\be
\frac{d^2 \phi}{d \theta^2} + 4\frac{\cos \theta}{\sin \theta} \cdot
\frac{d \phi}{d \theta}
+ \frac{\d V}{\d \phi} = 0 \; .
\label{sep12-14-10}
\ee
This corresponds to oscillations\footnote{For small curvature of the
potential at $\phi=\phi_+$, there is not enough ``time'' $\theta$ for
the oscillations to actually occur; the field just shifts from
$\phi_*$ towards $\phi_+$.}   near $\phi = \phi_+$, which are
damped at $\theta < \pi/2$, i.e.,
$t<0$.
The fact that the potential is no longer inverted is easy to understand:
normals to hypersurfaces $t/z = \mbox{const}$ are spacelike
for $z<|t|$ and timelike for $z> |t|$.

Even though we are interested in the solution at $t<0$,
we can pose a formal question of its behavior at positive $t$
($\theta > \pi/2$). Clearly, the solution is non-singular at
$\theta < \pi$, but generically
develops a singularity at
the null hypersurface  $\theta = \pi$, i.e., $t=+z$. There is nothing
wrong with that, since this null surface emanates from
the special ``point'' $t=0$, $z=0$ where the wall hits the adS boundary.

We note in passing that by fine tuning the potential
one can have solutions that are non-singular at $t=z$. Indeed, one can
arrange the potential in such
a way that the solution is symmetric under time reversal,
$t \to -t$, i.e., $\theta \to \pi - \theta$. In that case, there exists
also a domain wall at positive $t$ which expands to $z\to \infty$ as
$t \to +\infty$. Whether such solutions with fine tuned potentials
make sense from the CFT point of view remains to be understood.

To conclude, the (pseudo-)conformal scenario can be holographically
implemented in a theory of 5d scalar field whose potential has both
true and false vacua. In the 5d language, conformal rolling corresponds to
a spatially homogeneous transition from the false vacuum to the true one,
with a moving domain wall in between. We have demonstrated this within the
probe scalar field approximation, but since the solution is non-singular
in the relevant space-time domain, switching on gravity should not
change the picture, at least in the weak gravity regime.

\vspace{0.3cm}

S.S. thanks Riccardo Rattazzi for useful discussions.
The work of M.L. and V.R. has been supported by Russian Science Foundation
grant 14-12-01430.

\end{document}